\documentstyle[aps]{revtex}
\begin{document}
\title{Scheme for the preparation of the multi-particle entanglement in cavity QED}
\author{Guo-Ping Guo, Chuan-Feng Li\thanks{%
Electronic address: cfli@ustc.edu.cn }, Jian Li and Guang-Can Guo\thanks{%
Electronic address: gcguo@ustc.edu.cn }}
\address{Laboratory of Quantum Communication and Quantum Computation and Department\\
of Physics, University of Science and Technology of China, Hefei 230026,\\
People's Republic of China}
\maketitle

\begin{abstract}
Here we present a quantum electrodynamics (QED) model involving a
large-detuned single-mode cavity field and $n$ identical two-level atoms.
One of its applications for the preparation of the multi-particle states is
analyzed. In addition to the Greenberger-Horne-Zeilinger (GHZ) state, the W
class states can also be generated in this scheme. The further analysis for
the experiment of the model of $n=2$ case is also presented by considering
the possible three-atom collision.

PACS number(s): 03.65.Ud, 03.67.Lx, 42.50.Dv
\end{abstract}

Quantum entanglement, first noted by Einstein-Podolsky-Rosen (EPR)\cite{EPR}
and Schr\"{o}dinger\cite{s}, is one of the essential features of quantum
mechanics. Its famous embodiment $\Phi ^{\pm }=\frac 1{\sqrt{2}}(|11\rangle
\pm |00\rangle ),\Psi ^{\pm }=\frac 1{\sqrt{2}}(|10\rangle \pm |01\rangle )$
was shown by Bell\cite{Bell} to have stronger correlations than allowed by
any local hidden variable theory. For the multi-particle entanglement
states, there are many properties more peculiar than the two-party ones. For
example, the Greenberger-Horne-Zeilinger (GHZ) state\cite{ghz1,ghz2} $\Phi
^{ABC}=\frac 1{\sqrt{2}}(|111\rangle \pm |000\rangle ),$ a canonical
three-particle entanglement state exhibits the contradiction between local
hidden variable theories and quantum mechanics even for nonstatistical
predictions, as opposed to the statistical ones for the EPR states. Many
papers have discussed the multiparticle entanglement and its applications%
\cite{multi 1,multi 2}. In the paper\cite{w states}, the authors proved that
there exists another kind of peculiar genuine tripartite entanglement W
states $W=\frac 1{\sqrt{3}}(|001\rangle +|010\rangle +\left|
100\right\rangle ),$ which is inequivalent to the GHZ states in the sense
that they cannot been converted to each other even under stochastic local
operations and classical communication (SLOCC); that is, through LOCC but
without imposing that it has to be achieved with certainty\cite{slocc}. The
GHZ state is maximally in several senses\cite{w states 20}, for instance, it
maximally violates Bell-type inequalities, the mutual information of
measurement outcomes is maximal, it is maximally stable against (white)
noise and one can locally obtain from a GHZ state with unit probability an
EPR state shared between any two of the three parties. Another relevant
feature is that when any one of the three qubits is traced out, the
remaining two are in separable - and therefore unentangled - state. Thus,
the entanglement properties of the GHZ state are very fragile under particle
losses. Oppositely, the entanglement of the W state has the highest degree
of endurance against loss of one of the three qubits which is argued as an
important property in any situation where one of the three parties decide
not to cooperate with the other two\cite{w states}. For the generalized form 
$W_n=\frac 1{\sqrt{n}}\left| n-1,\text{ }1\right\rangle ,$ where $\left| n-1,%
\text{ }1\right\rangle $ denotes the totally symmetric state including $n-1$
zeros and $1$ ones, the concurrence (which is related to the formation
entanglement) of any reduced density operators $\rho _{k.u},$ $C_{k,u}(\rho
_{k.u})$ $=2/n,$ which indicates the maximal entanglement achievable for any
reduced two parties of system in any pure state\cite{w states,Distributed}.
The states which can been converted to each other under SLOCC belong to the
same class, then there are at least two inequivalent classes of
multiparticle entanglement states: the GHZ state class and the W state class%
\cite{w states}.

Recently, it has been realized that quantum resources can be useful in
information processing where quantum entanglement plays a key role in many
such applications like quantum teleportation\cite{telep}, computer\cite
{computer}, cryptography\cite{crypto}. And it has been shown that
multiparticle states have some advantages over the two-particle Bell states
in their application to cloning\cite{rendell 5,li}, teleportation\cite
{rendell 6}, and dense coding\cite{rendell 7}. Then the preparation and
manipulation of the entanglement states becomes a critical technique for
these quantum information processing. Many schemes such as these employing
the Jaynes-Cummings model in the cavity quantum electrodynamics (QED)\cite
{cat}, ion trap\cite{zheng 12}, NMR\cite{zheng 13} have been proposed. In
experiment, two particles entangled states have been realized in both cavity
QED\cite{qed} and ion traps\cite{ion}. But in most of the previous schemes
for quantum information processing in cavity QED and ion traps, the cavity
and ion motion act as memories. Thus the decoherence of the cavity field
becomes one of the main obstacles for the implementation of quantum
information in the cavity field, while in the ion traps is the difficulty to
achieve the joint ground state of the ion motion and the heating of the
ions. In paper\cite{zheng 16}, S$\phi $rensen and M$\phi $lmer have proposed
schemes for realizing quantum computation in the ion traps via virtual
vibrational excitations. They\cite{zheng 17} have also proposed a scheme for
the generation of multi-particle states in the GHZ state class in ion traps
without the requirement of the full control of the ion motion, which has
been accomplished in experiment\cite{e4}. In the cavity QED, Zheng and Guo%
\cite{zheng} have proposed a novel scheme for two-atom entanglement and
quantum information processing whose experiment implementation has also been
reported by Osnahgi {\it et al. }very quickly\cite{Haroche}{\it .}

In this paper, we present firstly a scheme for the generation of the
multi-particle entanglement states of both the GHZ state class and the W
state class in cavity QED. The scheme does not require the transfer of
quantum information between the atoms and the cavity. As the cavity is only
virtually excited, the requirement on the quality of the cavities is greatly
loosened and the efficient decoherence time of it is greatly prolonged.

Firstly, we consider the model of $n$ identical two-level atoms
simultaneously interacting with a single-mode cavity field. The interaction
Hamiltonian in the interaction picture is 
\begin{equation}
H_i=g\sum_{j=1}^n(e^{-i\delta t}a^{+}s_j^{-}+e^{i\delta t}as_j^{+}),
\end{equation}
where $s_j^{+}=\left| 1\right\rangle _{jj}\left\langle 0\right| $and $%
s_j^{-}=\left| 0\right\rangle _{jj}\left\langle 1\right| $, with $\left|
1\right\rangle _j$ and$\left| 0\right\rangle _j$ $(j=1,$ $2,$ $...,$ $n)$
being the excited and ground states of the $j$th atom, $a^{+}$ and $a^{-}$
are, respectively, the creation and annihilation operators for the cavity
mode, $g$ is the atom-cavity coupling strength, and $\delta $ is the
detuning between the atomic transition frequency $w_0$ and cavity frequency $%
w.$ In the case $\delta \gg g,$ there is no energy exchange between the
atomic system and the cavity. The effective Hamiltonian obtained by
adiabatically eliminating the atomic coherence is given by 
\begin{equation}
H=\lambda [\sum_{i,j=1}^n(s_j^{+}s_i^{-}aa^{+}-s_j^{-}s_i^{+}a^{+}a)],
\end{equation}
where $\lambda =g^2/\delta .$ This can been viewed as a generalized
Jaynes-Cummings model Hamiltonian describing a cavity mode interacting with $%
n$ atoms. When $n=1$, the Hamiltonian is 
\begin{equation}
H=\lambda (\left| 1\right\rangle \left\langle 1\right| aa^{+}-\left|
0\right\rangle \left\langle 0\right| a^{+}a),
\end{equation}
which represents the far-off-resonant case of the Jaynes-Cummings model\cite
{holland}. When $n=2,$ the Hamiltionian is 
\begin{equation}
H=\lambda [\sum\limits_{j=1,2}(\left| 1\right\rangle _{jj}\left\langle
1\right| aa^{+}-\left| 0\right\rangle _{jj}\left\langle 0\right|
a^{+}a)+(s_1^{+}s_2^{-}+s_1^{-}s_2^{+})],
\end{equation}
which has been shown to be useful in the generation of two-atom maximally
entangled states and the realization of quantum controlled-not gates and
quantum teleportation with dispersive cavity QED\cite{zheng}. The procedure
in this scheme, essentially insensitive to thermal fields and to photon
decay, opens promising perspectives for complex entanglement manipulations%
\cite{Haroche}.

Now we consider the case of multi-atom. Assume that the cavity field is
initially in the vacuum state, the Hamiltonian reduces to 
\begin{equation}
H=\lambda (\sum\limits_{j=1}^n\left| 1\right\rangle _{jj}\left\langle
1\right| +\sum_{i,j=1,i\neq j}^ns_j^{+}s_i^{-})
\end{equation}
It is obvious that there is no quantum information transfer between the
atoms and cavity. For the case of $n=3$, the Hamiltonian can be written as 
\begin{equation}
H=\lambda [\sum\limits_{j=1,2,3}\left| 1\right\rangle _{jj}\left\langle
1\right|
+(s_1^{+}s_2^{-}+s_1^{-}s_2^{+}+s_1^{+}s_3^{-}+s_1^{-}s_3^{+}+s_2^{+}s_3^{-}+s_2^{-}s_3^{+})].
\end{equation}
The first term describe the Stark shifts in the vacuum cavity, and the rest
terms describe the dipole coupling between any of the two atoms induced by
the cavity mode. Assume the atoms are initially in the state $\left|
001\right\rangle $, then the state evolution of the system can be
represented by 
\begin{equation}
W_3(t)=\frac{e^{-i3\lambda t}+2}3\left| 001\right\rangle +\frac{%
e^{-i3\lambda t}-1}3\left( \left| 010\right\rangle +\left| 100\right\rangle
\right) .
\end{equation}
With the choice of $\lambda t=\frac{2\pi }9,$ we obtain the W states\cite{w
states,wang} 
\begin{equation}
W_3=\frac 1{\sqrt{3}}(e^{i\frac{2\pi }3}\left| 001\right\rangle +\left|
010\right\rangle +\left| 100\right\rangle ),  \nonumber
\end{equation}
where the common phase factor $e^{-i\frac{5\pi }6}$ has been discarded.

Generally if initially the first $n-1$ atoms are in the state $\left|
0\right\rangle $ and the last atom is in $\left| 1\right\rangle ,$ the
evolution of the state goes as follows: 
\begin{equation}
W_n(t)=\frac{e^{-in\lambda t}+n-1}n\left| 0\right\rangle
_{_{1,2,...n-1}}\left| 1\right\rangle _n+\frac{e^{-in\lambda t}-1}n\left|
n-2,\text{ }1\right\rangle _{_{1,2,...n-1}}\left| 0\right\rangle _n,
\end{equation}
where $\left| n-2,\text{ }1\right\rangle _{_{1,2,...n-1}}$ denotes the
symmetric $n-1$ particles states involving $n-2$ zeroes and $1$ ones. With
the different choice of evolution time, one can get various $n$-particle
state of the W state class. This result can be understood from the
properties of the Hamiltonian: the parity bit of the state is unchanged in
the evolution process governed by this Hamiltonian, then the population
becomes distributed on all the state with the same parity bit which forms
the states of W state class. Obviously, only in the case of $n\leq 4$ can $%
\left| \frac{e^{-in\lambda t}+n-1}n\right| $ equal $\left| \frac{%
e^{-in\lambda t}-1}n\right| $ which represents the maximally entangled W
state. But if we measure the $n$th atoms at sometime $t$ and get $\left|
0\right\rangle _n$, then the rest $n-1$ atoms becomes in the state 
\begin{equation}
W_{n-1}=\frac 1{\sqrt{n-1}}\left| n-2,1\right\rangle .
\end{equation}
In this way, we can get the $(n-1)$-particle maximal entangled W states with
the probability of $\left| \frac{\sqrt{n-1}}n(e^{-in\lambda t}-1)\right| ^2$
which gets its maximal value for the case of $t=\frac \pi n$ and
approximately inversely proportionate to the atom number $n$.

Furthermore, using this generalized Jaynes-Cummings model, we can also
prepare the states in the GHZ class. Assume four atoms are initially in the
state $\left| 0011\right\rangle ,$ the evolution under the four-atom
Hamiltonian is 
\begin{eqnarray}
\left| \phi \right\rangle &=&\frac 16(e^{-i6\lambda t}+3e^{-i3\lambda
t}+2)\left| 0011\right\rangle +\frac 16(e^{-i6\lambda t}-3e^{-i3\lambda
t}+2)\left| 1100\right\rangle  \nonumber \\
&&+\frac 16(e^{-i6\lambda t}-1)(\left| 1001\right\rangle +\left|
0101\right\rangle +\left| 1010\right\rangle +\left| 0110\right\rangle ).
\end{eqnarray}
Also with the choice of $\lambda t=\frac \pi 3,$ we obtain a state belonging
to the GHZ state class 
\begin{equation}
\left| \phi \right\rangle =\frac{e^{-i\frac \pi 3}}2(\left|
0011\right\rangle +i\sqrt{3}\left| 1100\right\rangle ).
\end{equation}

Noticeably, although any $n$-particle W state can be generated
straightforwardly in the present scheme, the $m$-particle GHZ state where $%
m\geq 5$ cannot be prepared directly this way. But it has been well known
that entangled states involving higher numbers of particles can be generated
from entangled states involving lower numbers of particles by employing the
same procedure as entanglement swapping\cite{swapping}. The basic
ingredients are a Bell state measuring device and lower numbers of particles
entanglement states. Now it has been proved that there are at least two
inequivalent classes multi-particle entanglement states which can not been
converted to each other under SLOCC\cite{w states}. Then the lower numbers
of particles entanglement states for the preparation of a higher numbers of
particles state of GHZ state class or W state class must be GHZ state and W
state respectively\cite{Author}. Here we have shown that both classes of
states can be generated in the present scheme. Bell states measurement can
also be realized in this generalized Jaynes-Cummings model of the $n=2$ case%
\cite{zheng}. Then any multi-particle state of either the W state class or
the GHZ state class can be prepared in this scheme of the QED cavity.

The discussion on the experimental matters is similar to the paper\cite
{zheng}. The two atoms experiment to prepare EPR pair using the present
model of $n=2$ case has been realized recently\cite{Haroche}. As there is a
probability of $0.78,$ $0.19,$ $0.025$ respectively to have $0,$ $1,$ $2$
atoms in one atom pulse and events in which only one atom is detected in two
pulses are recorded, then in approximately $25\%$ of these events, there are
in fact two atoms in one of the pulses, one of them escaping detection. In
addition to the probabilities $P(e_1,g_2),$ $P(g_1,e_2),$ there are also
some spurious channels probabilities $P(e_1,e_2),$ $P(g_1,g_2)$ caused by
possible three atoms collision. All these probabilities could be calcualted
using the present multi-atom model in detail: 
\begin{eqnarray}
P(e_1,e_2) &=&P(g_1,g_2)=0.028(1-\cos (3\lambda t)),  \nonumber \\
P(e_1,g_2) &=&0.514+0.375\cos (2\lambda t)+0.111\cos (3\lambda t), \\
P(g_1,e_2) &=&0.430-0.375\cos (2\lambda t)-0.055\cos (3\lambda t),  \nonumber
\end{eqnarray}
where the two atom pulses are assumed initially in excited and ground state
respectively and the state discriminating errors are omitted. The result is
shown in Figure 1. The experiment of \cite{Haroche} have shown the existence
of $P(e_1,e_2)$ and $P(g_1,g_2)$, further experiment should reveal the
oscillation of these probabilities with the interacting time.

This generalized Jaynes-Cummings model requires the atoms be sent through
the cavity simultaneously, otherwise there will be an error. But the
influence of time difference is not as severe as expected. Even assume the
third atom in the excited state enters the cavity $10\%t_0$ later than the
other two ground state atoms ( the time difference between these two atoms
is nonsignificant ) in the generation of the W state. Then the three atoms
are finally prepared in the state $W_3(0.90t_0).$In the case that the third
atom leaves the cavity $10\%t_0$ earlier than the other two atoms, the three
atoms final state becomes 
\begin{equation}
W_3^{\prime }(0.90t_0)=\frac{e^{-i3\lambda t}+2}3\left| 001\right\rangle +%
\frac{e^{-i3\lambda t}-1}3e^{-i0.1\lambda t_0}\left( \left| 010\right\rangle
+\left| 100\right\rangle \right)
\end{equation}
If we still choosing $\lambda t_0=\frac{2\pi }9$ we have 
\begin{eqnarray}
\left| \langle W_3(0.90t_0)|W_3(t_0)\rangle \right| ^2 &\simeq &0.99, \\
\left| \langle W_3^{^{\prime }}(0.90t_0)|W_3(t_0)\rangle \right| ^2 &\simeq
&0.99.  \nonumber
\end{eqnarray}
The operation is only slightly affected.

In conclusion, we have present a generalized Jaynes-Cummings model involving
a single-mode cavity field and $n$ identical two-level atoms. One of its
applications for the preparations of the multi-particle states is analyzed.
In addition to the GHZ state, the W states can also be generated in this
scheme. The further analysis for the experiment of the model of $n=2$ case
is also presented by considering the possible three-atom collision. The most
distinct advantage of this model is that cavity initially is in vacuum state
and no quantum information transfer is required. Thus the requirement on the
quality factor of the cavity is greatly loosened and then implementation is
foreseeable.

This work was supported by the National Natural Science Foundation of China.

{\bf Figure Captions:}

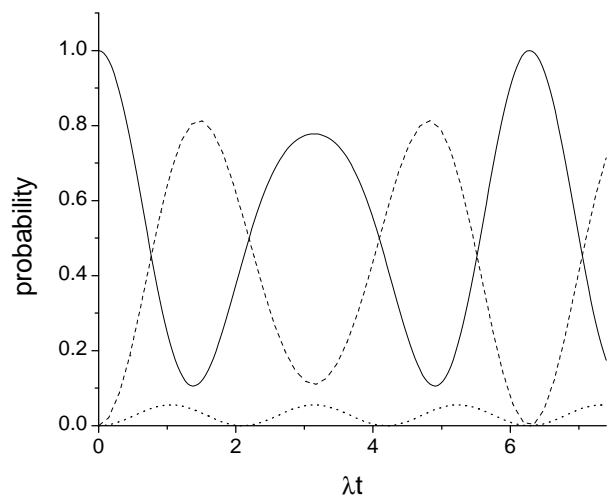
Figure 1: The graph of the measurement probabilities $P(e_1,g_2),$ $%
P(g_1,e_2),$ $P(e_1,e_2)$ and $P(g_1,g_2)$ versus $\lambda t.$ The solid
line denotes the probability $P(e_1,g_2),$ the dashed line denotes $%
P(g_1,e_2),$ and the dotted line denotes $P(e_1,e_2)$ and $P(g_1,g_2).$


\begin{references}
\bibitem{Author}  Using the same process as the paper\cite{swapping}, it can
be proved straightforwardly low numbers of W states must be employed for the
generation of any state in the W state class.
\end{references}
\end{document}